\documentclass[a4paper,12pt]{article}
\pdfoutput=1

\usepackage{soul}
\usepackage[usenames,dvipsnames]{color}
\usepackage{jheppub}
\usepackage{amsmath, amssymb, slashed, epsf, color, graphicx, latexsym}
\usepackage{epsfig}
\usepackage{graphics}
\usepackage{tikz}
\usepackage{graphicx,wrapfig,lipsum}
\usepackage{blindtext}

\newcommand{\be}{\begin{equation}}
\newcommand{\ee}{\end{equation}}
\newcommand{\bea}{\begin{eqnarray}}
\newcommand{\eea}{\end{eqnarray}}
\def\bse{\begin{subequations}}
\def\ese{\end{subequations}}
\newcommand{\IR}{\mathbb{R}} 
 
\def\IZ{\relax\ifmmode\hbox{Z\kern-.4em Z}\else{Z\kern-.4em Z}\fi}
\newcommand{\non}{\nonumber \\}

\def\del{{\partial}} 
\def\presub{\vspace{.5cm} \noindent}

\def\bi{\begin{itemize}} \def\ei{\end{itemize}}

\def\({\left(} \def\){\right)}

\newcommand{\nn}{\nonumber}

\newcommand{\p}{\partial}

\newcommand{\al}{\alpha}
\newcommand{\bt}{\beta}

\newcommand{\ep}{\epsilon}

\newcommand{\la}{\lambda}       

\def\laminf{\lambda_\infty}
\def\hh{\hat{h}}

\begin{document}
	
	\title{Triangle diagram, Distance Geometry \\ and Symmetries of Feynman Integrals}
	\author{Barak Kol and}
	\author{Subhajit Mazumdar}  

	\affiliation{Racah Institute of Physics, Hebrew University, Jerusalem 91904, Israel }
	\emailAdd{barak.kol, mazumdar.subhajit@mail.huji.ac.il}
	
	\abstract{We study the most general triangle diagram through the Symmetries of Feynman Integrals (SFI) approach. The SFI equation system is obtained and presented in a simple basis. The system is solved providing a novel derivation of an essentially known expression. We stress a description of the underlying geometry in terms of the Distance Geometry of a tetrahedron discussed by Davydychev-Delbourgo \cite{DavydychevDelbourgo1997}, a tetrahedron which is the dual on-shell diagram. In addition, the singular locus is identified and the diagram's value on the locus's two components is expressed as a linear combination of descendant bubble diagrams. The massless triangle and the associated magic connection are revisited.}
	
\maketitle

\section{Introduction}
 
Feynman Integrals are the computational core of Quantum Field Theory. Yet, despite over seventy years of work on their evaluation it appears that we do not have a general theory for it. The Symmetries of Feynman Integrals approach \cite{SFI} is a step in that direction. It considers a Feynman diagram of fixed topology (fixed graph), but varying kinematical invariants, masses and spacetime dimension. Each diagram is associated with a system of differential equations in this parameter space. The equation system defines a Lie group $G$ which acts on parameter space and foliates it into orbits. This geometry allows to reduce the diagram to its value at some convenient base point within the same orbit plus a line integral over simpler diagrams, namely with one edge contracted.

The SFI method is related to both the Integration By Parts method \cite{ChetyrkinTkachov1981} as well as to the Differential Equations method \cite{DE1:Kotikov1990, DE2:Remiddi1997, DE3:GehrmannRemiddi1999}, see also the textbooks \cite{SmirnovBook2006, SmirnovBook2012}. SFI novelties include the definitions of the group and its orbits, as well as the reduction to a line integral.

Other recent approaches to the evaluation of Feynman Integrals include a direct solution \cite{Kosower:2018obg}, avoiding squared propagators \cite{Bosma:2018mtf}, Intersection theory \cite{Mastrolia:2018uzb} and a development of loop-tree duality \cite{Runkel:2019zbm}.\footnote{see also \cite{Capatti:2019ypt, Aguilera-Verdugo:2019kbz}.}
 
\begin{figure}
\centering \noindent
\includegraphics[width=10cm]{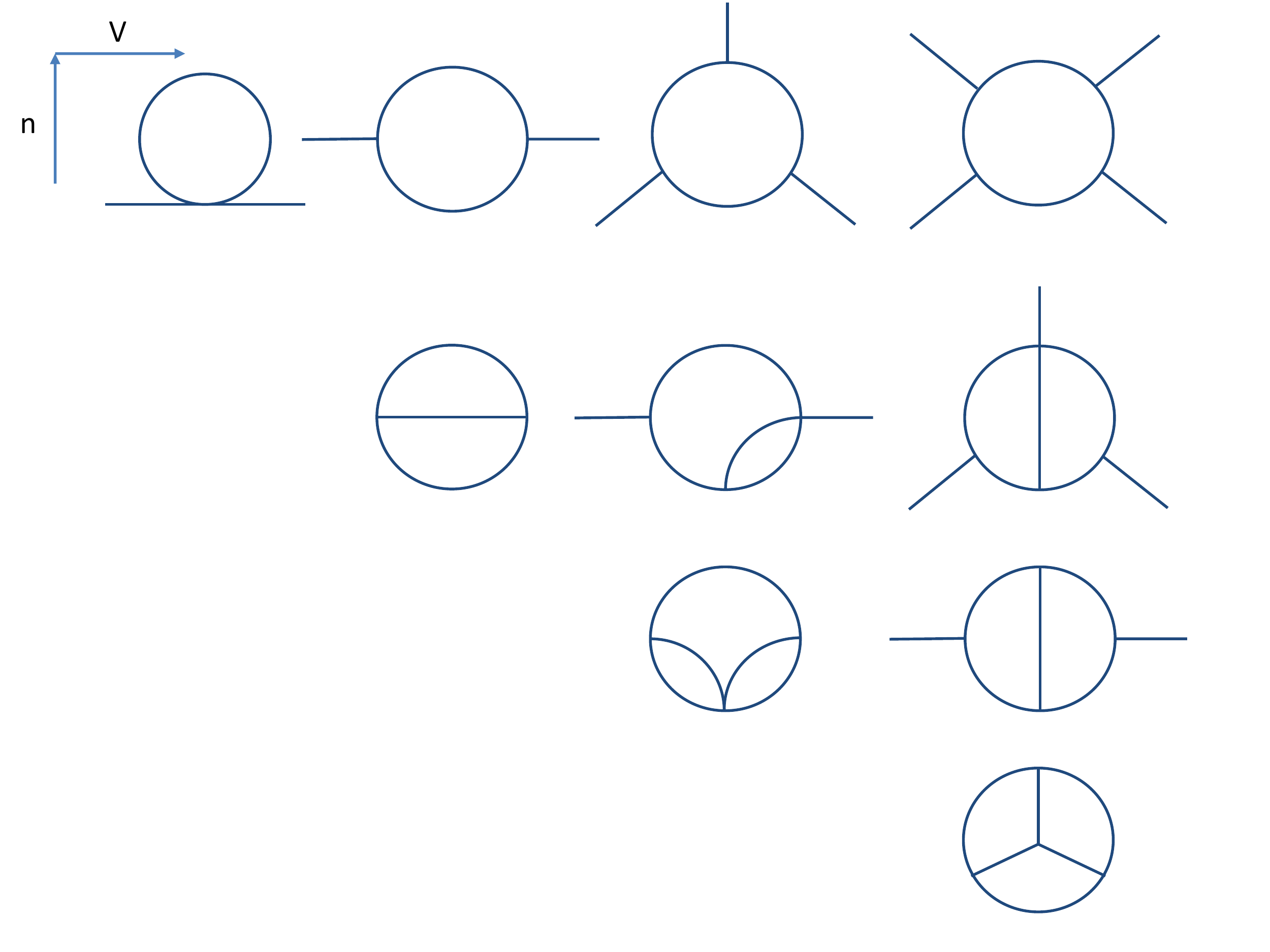} 
\caption[]{Roadmap to diagrams according to edge contraction.  Each column has diagrams of fixed number of vertices $V=1,2,3,4$. As contraction reduces $V$ by one the necessary sources for each diagram are always on its left. Each column in ordered according to the number of external legs $N$. The triangle is on the top row, second from the right.}
 \label{fig:DiagHierarchy} \end{figure}

SFI suggests to partially order all diagrams according to edge contraction as shown in fig. \ref{fig:DiagHierarchy} where the sources, or descendants, for each diagram are in the columns to its left. Several diagrams were already studied in this way: the bubble, diameter, vacuum seagull, propagator seagull and the kite \cite{bubble,diam,VacSeagull,PropSeagull,kite}. See also developments of the method in \cite{locus,minors}.

This paper studies the triangle diagram, namely the diagram with 3 legs and 1 loop. Clearly this is one of the simplest diagrams and its study through the SFI approach is intended to gain insight into both the diagram and the method.  We proceed to review the literature and to present the questions which we set to answer. 

Like all diagrams, the triangle can be considered in the plane of alpha (Schwinger) parameters. This representation was shown to offer a geometric interpretation of Feynman diagrams in terms of simplices  \cite{DavydychevDelbourgo1997} and the detailed application to the triangle was given in \cite{Davydychev2005}. The geometrical interpretation further suggested a decomposition of a general $n$-simplex into $n$ right handed simplices thereby recursively splitting the value of the $N=n$-point 1-loop diagram into a sum of $N!$ terms (sum decomposition) each having the same form, but depending on different sets of  $N-1$  variables  \cite{DavydychevDelbourgo1997,Davydychev2016}.  Another sum decomposition appeared from the functional relations of \cite{Tarasov2015}; when applied to the triangle these produce a decomposition of the massless triangle into a sum of 3 terms each having the same form, but depending on different sets of  2 variables \cite{Tarasov2019}. The similarity between these sum decompositions suggests that they are related. 

The Landau equations for the singularity locus of the triangle relate it to the planarity of the on-shell dual tetrahedron \cite{Landau1959}. Such a 3 particle singularity is known as an anomalous threshold, to distinguish it from the more common singularity of the bubble diagram, known as a normal threshold \cite{AnalyticS1966}.

The general massive triangle in 4d was evaluated in \cite{tHooftVeltman1978}  in terms of dilogarithms and improved in \cite{vanOldenborghVermaseren1989} including through the use of Gram determinants. It was expressed in terms of Appell function $F_3$ in \cite{Cabral-Rosetti_Sanchis-Lozano1998}. The maximally general case of general $d$  was expressed as a line integral in \cite{DavydychevDelbourgo1997}, evaluated in terms of the Appell function $F_1$ through dimensional recursion relations in \cite{Tarasov2000} and somewhat improved in \cite{FleischerJegerlehnerTarasov2003}, see also \cite{Davydychev:2001uj}. Finally we note that IBP relations for the triangle diagram were discussed in \cite{Bern:2017gdk}, where consequences of dual conformal symmetry were studied.

The massless triangle satisfies the so-called magic connection. It depends on 3 kinematic invariants with an $S_3$ permutation symmetry and the same is true of the diameter diagram (two-loop vacuum) which depends on 3 masses. Surprisingly the two essentially coincide once the spacetime dimension is transformed  \cite{magic1995}.  This relation was called the magic connection.

\presub We shall be interested in the following questions  \bi
\item What is the SFI equation system for the triangle? 
\item What is the geometry in parameter space including orbit co-dimension and singular locus?
\item Can the system be solved through SFI on the singular locus? In general?
\item Does SFI shed light on the sum decomposition? on the magic connection?
\ei

The paper is organized as follows. We start in section \ref{sec:setup} by setting up the definitions, presenting useful facts on tetrahedron geometry and an account of the alpha parameters presentation together with the associated sum decomposition. In section \ref{sec:eq_sys} the equation system and the associated SFI group $G$ are presented,  
followed by a study the geometry of parameter space. Section \ref{sec:solns} describes the solutions to the equation system: first the reduction of the integral at the two components of the singular locus and then the general solution is derived. The massless triangle and the magic connection are discussed in section \ref{sec:magic} from an SFI perspective.  Finally, section \ref{sec:summ} is a summary and discussion. An Appendix contains a generalization of the tetrahedron geometry to higher dimensional simplices.   

\section{Set-up}
\label{sec:setup}

\subsection{Definitions}
\label{sec:def}

\begin{figure}[ht]
\begin{center}
\begin{tikzpicture}[scale=.6, transform shape]
     \draw [black,thick,domain=5.19:7.2] plot ({0}, {\x}); 
 \draw[black, thick] (3,0) -- (0,5.19);
     \draw[black, thick] (-3,0) -- (0,5.19);
      \draw[black, thick] (-3,0) -- (3,0);
     \draw[black, thick] (-5,-1.3) -- (-3,0);
      \draw[black, thick] (5,-1.3) -- (3,0);
      \node at (-0.5,7.0) {$p_{1}$};
     \node at (-4.5,-1.4) {$p_{3}$};
     \node at (4.5,-1.4) {$p_{2}$};
     \node at (0,2.0) {$l$};
     \node at (0.0,-1.0) {$x_{1}$};
        \node at (2.5,2.5) {$x_{3}$};
         \node at (-2.5,2.5) {$x_{2}$};
          \node at (3.5,0.2) {$b$};
          \node at (-3.5,0.2) {$c$};
          \node at (0.3,5.40) {$a$};
     \draw [black,thick,-> ,domain=358:30] plot ({1.2*cos(\x)}, {2+1.2*sin(\x)});
\end{tikzpicture}
\caption{The triangle diagram. $p_1,\, p_2,\, p_3$ are the external currents of energy-momentum while $x_1,\, x_2,\, x_3$ are the squared masses of the respective propagators ($x_1 \equiv m_1^{~2}$, etc.). The vertices are denoted by $a,\, b,\, c$.}
\label{fig:tri}
  \end{center}
\end{figure}
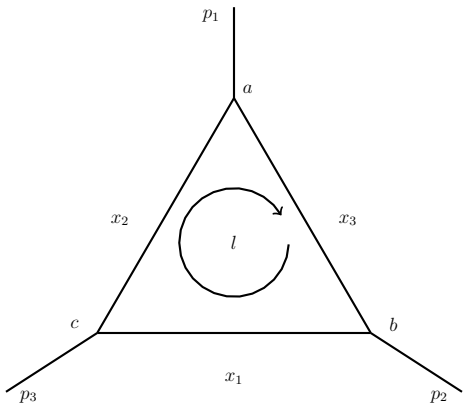

The subject of this paper is the triangle diagram shown in fig. \ref{fig:tri} and the associated Feynman integral defined by \be
 I = \int \frac{d^dl}{\prod_{i=1}^3 \( k_i^{~2}-m_i^{~2} \)}~.
 \label{def:I}
 \ee
 The propagator currents can be chosen as \footnote{
 An alternative practical choice is given by $k_1=l,\, k_2=l+p_3,\, k_3=l-p_2$.}
 \be
 k_i = l + \( p_{i+1} - p_{i-1} \)/ 3 ~~, i=1,2,3.
 \ee

The most general parameters which define the diagram are shown in the figure. As usual, the masses appear in the integral only through their squares. In order to parameterize the kinematical invariants, given there are 2 independent external momenta and thus 3 independent kinematical invariants, we choose to parameterize them symmetrically through $p_1^{~2},\, p_2^{~2}$ and $p_3^{~2}$. Altogether, 
the parameter space $X$ is given by \be
X= \left\{ \(x_1,x_2, x_3, x_4, x_5, x_6 \) = \( m_1^{~2},\, m_2^{~2},\, m_3^{~2},\, p_1^{~2},\, p_2^{~2},\, p_3^{~2}  \) \right \}~.
\label{def:X}
\ee
 We consider a general spacetime dimension $d$ where the mass dimension of the integral is $d - 6$.
 
The diagram and the associated integral are invariant under a  $\Gamma=S_3$ permutation symmetry. Its elements consist of rotations, which are generated by a third of a full rotation, and of 3 reflections through appropriate axes.

\presub {\bf Tetrahedron volume}. As a preliminary we define two quantities which describe the volumes of general triangles and tetrahedra and will be useful throughout the paper. The Heron / K\"all\'en invariant $\la$ is defined for any three quantities $x,y,\, z$ by
\be
\la \( x,y,z \) := x^2 + y^2 + z^2 - 2\, x\, y -2\, x\, z - 2\, y\, z ~.
\label{def:la}
\ee
If $x,y,z$ denote the squared lengths of the sides of a triangle, then its squared area  is given by $-\la/16$, see e.g. \cite{SFI,bubble} and references therein. Any trivalent vertex $v$ defines a corresponding $\la$  \be
\la_v:=\la(x_i,x_j,x_k)
\label{def:lav}
\ee 
where $i,j,k$ are the edges incident on $v$. In particular, the external momenta of the triangle are incident on the $\infty$-vertex and form a triangle whose Heron / K\"all\'en invariant is denoted by \be 
\la_\infty := \la\(x_4, x_5, x_6 \) ~.
\label{def:la_infty}
\ee

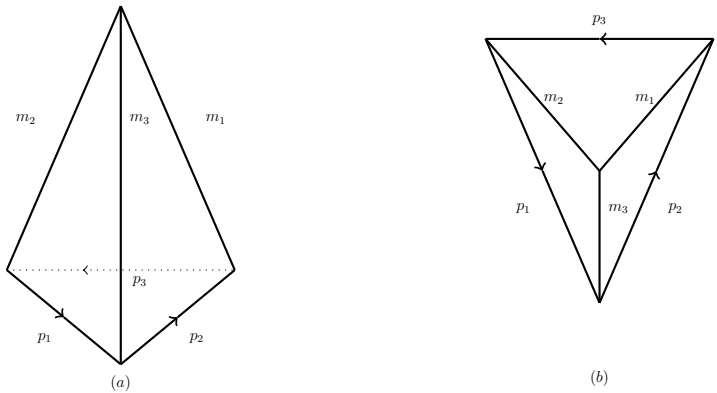
\begin{figure}[ht]
\begin{center}
\begin{minipage}{.2\textwidth}
\begin{tikzpicture}[scale=.5, transform shape]
\draw [black,thick,domain=-2.5:7] plot ({0}, {\x}); 
 
      \draw[black, dotted] (-3,0) -- (-1,0);
      \draw[black, dotted,<-] (-1,0) -- (3,0);
      
      \draw[black, thick,->] (0,-2.5) -- (1.5,-1.25);
      \draw[black, thick] (1.5,-1.25)--(3,0) ;
     \draw[black, thick,->] (-3,0) -- (-1.5,-1.25);
     \draw[black, thick] (-1.5,-1.25)--(0,-2.5);
      \draw[black, thick] (3,0) -- (0,7);
     \draw[black, thick] (-3,0) -- (0,7);
      \node at (0.5,-0.3) {$p_3$};
      \node at (2.0,-1.8) {$p_2$};
      \node at (-2.0,-1.8) {$p_1$};
      \node at (0.5,4) {$m_3$};
       \node at (2.5,4) {$m_1$};
        \node at (-2.5,4) {$m_2$};
        \node at (0,-3) {$(a)$};
        \end{tikzpicture}
\end{minipage}
\hspace{3cm}
\begin{minipage}{.2\textwidth}
\begin{tikzpicture}[scale=.5, transform shape]
  \draw [black,thick,domain=0.5:-3] plot ({0}, {\x}); 
 \draw[black, thick] (3,4) -- (0,0.5);
     \draw[black, thick] (-3,4) -- (0,0.5);
      \draw[black, thick] (-3,4) -- (0,4);
      \draw[black, thick,->] (3,4)--(0,4) ;
      \draw[black, thick,->] (0,-3)-- (1.5,0.5);
      \draw[black, thick] (1.5,0.5)--(3,4) ;
     \draw[black, thick,->] (-3,4)--(-1.5,0.5) ;
      \draw[black, thick] (0,-3)--(-1.5,0.5) ;
      \node at (-2.0,-0.5) {$p_1$};
      \node at (1.2,2.4) {$m_1$};
      \node at (-1.2,2.4) {$m_2$};
      \node at (0.5,-0.5) {$m_3$};
       \node at (2.0,-0.5) {$p_2$};
        \node at (0.0,4.5) {$p_3$};
        \node at (0,-5) {$(b)$};
\end{tikzpicture}
\end{minipage}
\caption{The relevant tetrahedron with sides of squared-length $x_1, \dots, x_6$. It is dual to the vacuum closure of the triangle diagram. (a) A 3d image, (b) a plane projection. This tetrahedron appears also in the Landau equations \cite{Landau1959}.}
\label{fig:tetra}
  \end{center}
\end{figure}

$B_{3}$ will denote a cubic polynomial given by
\begin{eqnarray}
B_3
&=&{x_1}^2 x_4+x_1 {x_4}^2+{x_2}^2 {x_5}+{x_2} {x_5}^2+{x_3}^2 {x_6}+{x_3} {x_6}^2 \non 
&+& x_1 x_2 x_6+x_1 x_3 x_5+x_2 x_3 x_4+x_4 x_5 x_6 \\ 
&-&(x_2 x_5 (x_1 + x_3 + x_4 + x_6)+x_3 x_6 (x_1 + x_2 + x_4 + x_5) + x_1 x_4 (x_2 + x_3 + x_5 + x_6))\nonumber ~.
\label{def:B3}
\end{eqnarray} 
$(-B_3)/144$ expresses the squared volume of a tetrahedron in terms of the squared lengths of its sides (the edges associated with $x_1,x_2,x_3$ meet at a point, while those associated with $x_4,x_5,x_6$ form a triangle). This is known as Tartaglia's formula, after the Italian mathematician-engineer (1499/1500-1577) who published it, yet essentially it was already known to the Italian painter Piero della Francesca (c. 1415-1492) \cite{math_tetrahedron,wiki_tetrahedron,tartaglia,wiki_tartaglia,francesca,wiki_francesca}.  $B_3$ appeared in the physics literature in the work of Baikov on the 3-loop vacuum diagram (the tetrahedron) \cite{Baikov1996a, Baikov1996b}. Therefore we shall refer to $B_3$ as the Tartaglia / Baikov polynomial. The tetrahedron relevant to the triangle diagram is shown in fig. \ref{fig:tetra}.

We noticed the following useful relations among these quantities \bea
	\(\del^1 B_3\)^2- 4\, x_4\, B_3 &=& \la_a\, \laminf \label{id1} \\
	\( \del^1 + \del^2 + \del^3 \) B_3 &=& \laminf \label{id2}   ~.
\eea
$\la_a=\la(x_2, x_3, x_4)$ is the $\la$ variable (\ref{def:lav}) associated with the vertex $a$ in fig. \ref{fig:tri}. The first line, (\ref{id1}), can be permuted cyclically to produce two additional identities. Moreover, for a Euclidean tetrahedron the derivative $\del^1 B_3$ is given by \be
 \del^1 B_3 = \sqrt{\la_a \laminf} \cos (\alpha)
 \label{id3}
 \ee
 where here $\alpha$ denotes the angle between the $a$ and $\infty$ faces of the dual tetrahedron. The identities appeared already in equations (9,10) of \cite{FleischerJegerlehnerTarasov2003}, while identity (\ref{id3}) was noted already in theorem 1 of \cite{Khimsh_etal2015}.
 
The generalization of both $\la$ and $B_3$ to a simplex of arbitrary dimension is given by the Cayley -- Menger determinant \cite{Cayley1841,Menger1928}\footnote{
Introduced by Cayley in 1841 for the 4-simplex and straightforwardly generalized to arbitrary dimensions by Menger in 1928 as part of developing an axiomatic approach to geometry.},  
 which applies also to pseudo-Riemannian metrics. The identities (\ref{id1},\ref{id2}) are generalized to an arbitrary $n$-simplex in appendix \ref{sec:n_simplex} where they are proven and some geometrical interpretation is provided. 

\subsection{Alpha plane and sum decomposition}
\label{subsec:alpha}

In terms of alpha (Schwinger) parameters the triangle integral is given by \be
 I = c_{\Delta}\, \int_{\Delta_\bt} V^{\frac{d-6}{2}}
\label{I_beta}
\ee
 where the triangle constant is given by \be
 c_{\Delta} := -i \pi^{\frac{d}{2}}\, \Gamma \(\frac{6-d}{2}\) ~;
\label{def:cTr}
 \ee
 the beta integration is over \be
 \int_{\Delta_\bt} := \int_0^1 d\bt^1\, d\bt^2\, d\bt^3\, \delta \(\bt_1+\bt_2+\bt_3-1\) ~,
 \ee
namely, the two dimensional simplex (a triangle with vertices at $u_1=(1,0,0),\, u_2=(0,1,0)$ and $u_3=(0,0,1)$); finally the standard Kirchhoff-Symanzik polynomial $V$ is given by 
\be
	V(\{\bt^i \}_{i=1}^3;\,\{x_j\}_{j=1}^6) := x_1\, \bt^1 + x_2\, \bt^2 + x_3\, \bt^3 - \( x_4\, \bt^2\, \bt^3 + x_5\, \bt^3\, \bt^1+  x_6\, \bt^1\, \bt^2 \) ~.
	\label{def:V}
\ee

\presub {\bf Change of variables and geometric interpretation}. The following is closely related to the results of \cite{DavydychevDelbourgo1997} regarding geometrical interpretation and sum decomposition (split). After the forthcoming description we shall comment on this relation. 

Given that $V$ is quadratic in the $\beta_i$ variables, we can map them into 2d $q$ variables such that $V$ is in the form $V=V_0 + q^2$ where $q^2$ is a canonical quadratic form, namely, it is diagonal and its entries all belong to $\{1,-,1,0\}$. We denote by $O$ the point where the extremum of $V$ is obtained. Its $\beta$ coordinates are given by $\beta_1(O)=\del_1 B_3/\la_\infty$ where $\laminf, B_3$ were defined in (\ref{def:la_infty},\ref{def:B3}), and similarly for  $\beta_2(O),\, \bt_3(O)$.

We notice that length squared of the edge suspended from $u_2$ to $u_3$, computed in the $q$ coordinates, is $x_4 \equiv p_1^{~2}$, and similarly for the other two edges. This allows us to identify the integration region in the $q$ coordinates with the triangle in momentum space formed by the external momenta $p_1,\, p_2,\, p_3$. Accordingly, the integration measure is transformed into \be
 \int_{\Delta_\bt} \to \frac{1}{\sqrt{|\la_\infty|/4}}  \int_{\Delta_q} d^2q
\label{def:Deltaq}
 \ee
 where $\Delta_q$ denotes the triangle in the $q$ variables.

We can furthermore enhance this geometrical picture to incorporate $V_0$ as well. The 2d $q$ plane contains a marked point $O$. From it erect an abstract 3rd axis perpendicular to the $q$ plane, and mark a point $\widehat{O}$ such that if we denote $\vec{h}=\overrightarrow{O \widehat{O}}$ then its length squared is given by $h^2 = V_0$  where $V_0$ is the extremal value of $V$ (this defines the signature of this extra dimension, namely whether it is spacelike or timelike). Now $V$ is given by \be
V(q) = V_0 + q^2 = {\overrightarrow{O \widehat{O}}}^2 + {\overrightarrow{Oq}}^2 \equiv  {\overrightarrow{\widehat{O} q}}^2
 \label{def:Vq}
\ee
where \be
V_0 \equiv {\overrightarrow{O \widehat{O}}}^2 \equiv h^2 :=  \frac{B_3}{\la_\infty} ~,
 \label{def:V0_and_h}
\ee
and ${\overrightarrow{Oq}}^2$ is in canonical form. This means that $V(q)$ is interpreted as the squared-distance of $q$ from $\widehat{O}$.

Altogether, the $q$ variables transform the Schwinger plane expression (\ref{I_beta}) through (\ref{def:Deltaq}) into \be
  I(x) = \frac{c_\Delta}{\sqrt{|\laminf|/4}} \int_{\Delta_q} d^2 q\, V^{\frac{d-6}{2}} 
\label{gen_express}
 \ee
where $c_\Delta$ is defined in (\ref{def:cTr}) and $V$ in (\ref{def:Vq}) . 
In terms of the tetrahedron shown in fig. \ref{fig:tetra}. The integration is over the interior of the triangle formed by $p_1,\, p_2,\, p_3$; the integrand is defined through the distance to the vertex where $m_1, m_2$ and $m_3$ meet. This tetrahedron is dual to the graph of the Feynman diagram, fig.  \ref{fig:tri}, and it is on-shell in the sense that the edge lengths are given by the masses of the corresponding (dual) propagators. 

\begin{figure}[ht]
\begin{center}
\begin{tikzpicture}[scale=.8, transform shape]
 
  \draw[black, thick] (6,0) -- (0,5.19);
     \draw[black, thick] (-2,0) -- (0,5.19);
      \draw[black, thick] (-2,0) -- (6,0);
          
     \node at (0.5,5.5) {$(1,0,0)$};
      \node at (7.2,0.2) {$(0,1,0)$};
      \node at (-2.9,0.2) {$(0,0,1)$};
       \node at (1.5,-0.5) {$a_{1}$};
        \node at (3.3,3.0) {$a_{3}$};
         \node at (-1.8,2.8) {$a_{2}$};
                  
\node at (0.3,3.0) {$c_{1}$};
        \node at (2.9,1.2) {$c_{2}$};
         \node at (-0.3,1.2) {$c_{3}$};         
         
     \node at (0.8,1.9) {$O$};
       \draw[black, dotted] (-2,0) -- (1.5,1.3);
        \draw[black, dotted] (0,5.19) -- (1.5,1.3);
        \draw[black, dotted] (6,0) -- (1.5,1.3);
        \draw[black, thick] (1.5,0) -- (1.5,1.3);
        \draw[black, thick] (2.7,2.9) -- (1.5,1.3);
        \draw[black, thick] (-1,2.5) -- (1.5,1.3);
\end{tikzpicture}
\caption{Splitting the integration region into 6 right handed triangles. The figure assumes a Euclidean (spacelike) plane, but the procedure applies to a general signature.}
\label{fig:triangle_1}
  \end{center}
\end{figure}
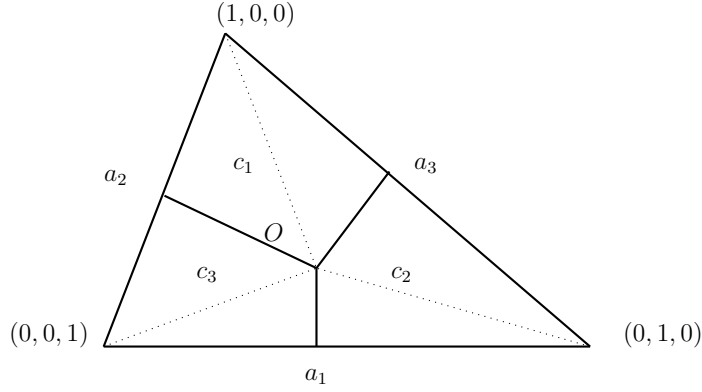

\presub {\bf Sum decomposition}. The metric in $q$ space enables a natural decomposition of the triangular integration region $\Delta_q$ thereby leading to a sum decomposition of the integral (\ref{def:I}). As shown in fig. (\ref{fig:triangle_1}), the triangular domain is first divided into 3 triangles by connecting the point $O$ with the 3 vertices. The squared-length of the segment $Oa$ connecting $O$ to  vertex $a$ where $p_2$ and $p_3$ meet is denoted by $c_1^{~2}$ and is given by \be
c_1^{~2} := m_1^2 - {\overrightarrow{O \widehat{O}}}^2 = x_1 -\frac{B_3}{\la_\infty}~.
 \label{def:c}
 \ee
Similarly we define and express $c_2^{~2},\, c_3^{~2}$.

Next each triangle is bisected by a height from $O$ to an opposite edge. We denote the distance squared to edge $m_1$ by $a_1^{~2}$ and it is given by
\bea
a_1^{~2} &=& c_2^{~2}-\frac{(c_2 \cdot p_1)^{2}}{p_1^{~2}} \equiv c_3^{~2}- \frac{(c_3 \cdot p_1)^{2}}{p_1^{~2}}  \non
	&=& -\frac{(\del_1 B_3)^2}{4 x_4 \la_\infty} = - \frac{\la_a}{4 x_4} - \frac{B_3}{\la_\infty}
\label{def:a}
\eea
where the last equality uses the identity (\ref{id1}) and $\la_a:=\la(x_2,x_3,x_4)$ is the Heron / K\"all\'en invariant associated with vertex $a$. Similarly we define and express $a_2^{~2},\, a_3^{~2}$.

In this way the triangular integration region is split into 6 right-handed triangles, each one including the point $O$ as a vertex. Correspondingly the triangle integral (\ref{def:I}) decomposes into a sum of 6 terms \be
 I = \frac{c_\Delta}{\sqrt{|\la_\infty|/4}} \left[ F(h^2, c_1^{~2}, a_2^{~2}) + F(h^2, c_1^{~2}, a_3^{~2}) + cyc. \right]
 \label{IsumF}
 \ee
   where the function $F$ is defined by  
\bea
	F(h^2,\, c^2,\, a^2) &:=&  \int_{\Delta_{a,c}} d^2q \,  \(h^2+q^2\)^\frac{d-6}{2} \non
	\int_{\Delta_{a,c}}  d^2q \, &:=& \int_0^{|a|} dq_y \int_0^{\frac{|b|}{|a|} q_y} dq_x 
 \label{def:F}
\eea
and $|a|:=\sqrt{|a^2|},\, |b|:=\sqrt{|c^2-a^2|}$ are the side-lengths of a right angle triangle. $q^2$ is the squared-length of the vector $\overrightarrow{Oq}$. For a spacelike $q$ plane $q^2=-q_x^{~2}-q_y^{~2}$, while for a $1+1$ signature one of these signs should be changed.
Note that the integrand of $F$ is essentially that of the full triangle integral (\ref{I_beta},\ref{def:Vq}) and only the integration region is restricted to a right handed triangle. In this way each $F$ summand depends on only 3 dimensionful parameters out of 6.

\presub {\bf Comments}.

$4d$ value. The $\ep$ expansion around $4d$ is known to be expressed in terms of the dilogarithmic function \cite{tHooftVeltman1978}.  In fact the $d \to 4$ limit of (\ref{def:F}) is finite. It depends on the signatures of $q^2$ and $h^2$. For some choice of signature we were able to evaluate the integrals and we obtained 
\bea
4 \int_{0}^{a} dt\int_{0}^{\frac{b }{a}t}dx\frac{1}{h^2+t^2-x^2}&=&\nonumber  -Li_{2}\left(-\frac{a^2}{h^2}\right)+Li_{2}\left(\frac{b^2-a^2}{h^2}\right) \non
&+&2\bigg(Li_{2}\left(\frac{a}{a-\sqrt{a^2+h^2}}\right)+Li_{2}\left(\frac{a}{a+\sqrt{a^2+h^2}}\right) \non
&-&Li_{2}\left(\frac{a+b}{a-\sqrt{a^2+h^2}}\right)-Li_{2}\left(\frac{a+b}{a+\sqrt{a^2+h^2}}\right)\bigg) \non
\label{dilog}
\eea

Generalization to $N$-point functions. The identification of the integration simplex in $\bt$ Schwinger parameters with the dual on-shell simplex generalizes to 1-loop diagrams with any number of external legs $N$.

Relation with \cite{DavydychevDelbourgo1997}.  The results of this section are closely related to those of \cite{DavydychevDelbourgo1997,Davydychev2016}, including the geometric interpretation and the sum decomposition. However, the integration domain is somewhat different, being the $N-1$ simplex rather than the corresponding hypersphere, and some readers may find the current presentation to be clearer.

Relation with Appell functions.  We compared the expression for the triangle (\ref{IsumF}-\ref{def:F}) with the expression in terms of Appell functions in \cite{FleischerJegerlehnerTarasov2003}, equations (74-81), and found them to coincide through numerical evaluation at numerous randomly selected parameter values.  


\section{Equation system}
\label{sec:eq_sys}

We will study the diagram through the Symmetries of Feynman Integrals method (SFI) described in \cite{SFI}. Briefly, one varies the integral with respect to infinitesimal re-definitions of loop momenta thereby giving rise to a set of differential equations which the integral satisfies in parameter space $X$.  Let us determine the equation set for $I$ and the associated group $G$.

\presub {\bf The SFI group}. Let us choose the 3 independent currents to be $\{q_m \}_{m=1}^3 = \{l,p_2,p_3\}$. The irreducible numerators (or irreducible scalar products) are defined to be the quotient of current quadratics by current squares (or propagators)  \be 
	Num = Q/S =  {\rm Sp} \{ q_m \cdot q_n \}_{m,n=1,2,3} / {\rm Sp} \{k_i^{~2},\,p_i^{~2}\}_{i=1,2,3} = \{ 0 \} ~.
\label{num}
\ee
This means that there are no irreducible numerators and hence all the 7 current variation operators in the upper triangular group $T_{1,2}$ define differential equations for $I$ and therefore belong to the SFI group $G$ \be
	 G = T_{1,2} = {\rm Sp} \{ l \del_l,\, p_i \del_l,\, p_i \del_{p_j} \}_{i,j=2,3} ~.
\label{G_SFI}
 \ee

\begin{figure}[ht]
\begin{center}
\begin{tikzpicture}[scale=.4, transform shape]
   \draw[black, thick] (4.5,0) -- (5.5,0);
     \draw[black, thick] (-5.5,0) -- (-3.5,0);
      \draw[black, thick] (2.5,0) -- (4.5,0);
     \draw[black, thick] (-3.5,0) -- (-2.5,0);     
        \draw [thick](0,0) circle (2.5cm);
 \end{tikzpicture}
\caption{The sources (or descendants) for the triangle consist of three possible bubble diagrams (see \cite{bubble}), corresponding to the omission of either propagator $i=1,2$ or $3$.}
\label{fig:sources}
  \end{center}
\end{figure}
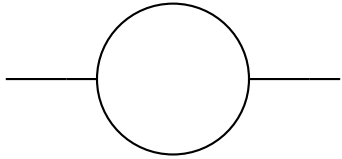

\presub {\bf The SFI equation system}. The SFI equation system consists of 7 equations which correspond to the operators in $G$. We choose a basis which is compatible with the $S_3$ symmetry group and has the shortest possible source terms. We find that $p_2 \del_{p_2}$ produces one such equation, namely \be
 0 = I + 2 s^1_b \frac{\del}{\del x_3} I+ 2 s^6_\infty \frac{\del}{\del x_4} I + 2 x_5 \frac{\del}{\del x_5} I -  \frac{\del}{\del x_3} O_1 I 
\label{SFIeq}
\ee 
 where for any trivalent vertex $v$, with incident vertices $i,j,k$, the $s^i_v$ variables denote \be
	s^i_v := \(x_j + x_k-x_i \)/2 \equiv -\frac{\del \la_v}{4\, \del x_i}
 \label{def:s}
 \ee
and $O_i$ denotes the operation of omission (or contraction) of propagator $i$. Such a contraction of the triangle produces the bubble topology shown in fig. \ref{fig:sources}. 

5 other equations are gotten by $S_3$ permutations. The seventh and last equation is the dimension equation \be
	 0 = (d-6) I - 2 \sum_{i=1}^6 \frac{\del}{\del x_i} I 
 \label{dimeq}
\ee
which is generated by $\sum_{i=1}^3 q_i \del_{q_i}$ and is equivalent to performing a dimensional analysis.

The equation system thus obtained can be summarized in matrix form by \be
 0 = c^a\,  I - 2\, \(Tx\)^a_j\, \partial^j\, I + J^a , ~ a=1,\dots, 7
 \label{eq_set}
\ee
where the generator matrix is given by \be
\(Tx\)^a_j = \( \begin{array}{cccccc}
0	& s_c^1 		& 0 & s_\infty^5 & 0 	& x_6 \\
 0	& 0	& s_a^2 	& x_4 &  s_\infty^6 & 0 \\
  s_{b}^3 	& 0 		& 0	& 0 & x_5 	& s_\infty^4 \\
0 & 0 	& s_b^1	& s_\infty^6 		& x_5	& 0 \\
 s_c^2	& 0	& 0	& 0		& s_\infty^4	& x_6 \\
 0 	& s_a^3		& 0 & x_4 & 0 	& s_\infty^5 \\
  x_1	 & x_2	& x_3	& x_4		& x_5 	& x_6 \\
\end{array} \) ~;
\label{def:Tx}
\ee
the $x_i$-independent constants are given by  \be
c^a= \left(
\begin{array}{c}
-1 \\
-1 \\ 
-1 \\
-1 \\
-1 \\
-1 \\
d-6 \\
\end{array}
\right) ~;
\label{def:c^a}
\ee
and finally, the sources are given by \be
J^a= \left(
\begin{array}{c}
 \frac{\del }{\del x_2} O_1\, I \\
 \frac{\del }{\del x_3} O_2\, I \\
 \frac{\del }{\del x_1} O_3\, I \\
 \frac{\del }{\del x_3} O_1\, I \\
 \frac{\del }{\del x_1} O_2\, I \\
 \frac{\del }{\del x_2} O_3\, I \\
0 \\
\end{array}
\right) ~.
\label{def:J}
\ee

This equation system was confirmed to hold for the expression in $\al$ space (\ref{I_beta}-\ref{def:Deltaq}) and against the program FIRE \cite{Smirnov:2014hma}. Its determination answers the first question posed in the introduction section.

\presub {\bf Source simplification}. We can use the SFI equations for the bubble diagram, see \cite{bubble}, to eliminate derivatives of the bubble which appear in the source term (\ref{def:J}), in favor of the bubble and its tadpole sources.  Denoting by  $I_3 \equiv O_3\, I$, a bubble with squared-masses $x_1,\, x_2$ we have \be
 \frac{\la}{2}\, \frac{\del}{\del x_1} I_3 = -(d-3) s_c^1\, I_3 + x_2\, j_2' - s_c^6\, j_1'
\label{source_simplf}
 \ee
 where the tadpoles are described by $j_i=j(x_i),\, j(x)= c_t\, x^{(d-2)/2},\, c_t=-i \pi^{d/2}\, \Gamma \( \frac{2-d}{2} \)$ and the bubble $s$ variables are defined in (\ref{def:s}). Analogous relations are gotten by permutations.

\subsection{Geometry of parameter space}
\label{sec:geom}

In this section we analyze the geometry in parameter space.

\presub {\bf $G$-orbit co-dimension and 6-minors}. The equation system (\ref{eq_set}) consists of 7 equations in a 6 dimensional parameter space. The dimension of the $G$-orbit through any point $x \in X$ is given by the rank of $Tx$ at that point. 

In order to determine the rank we follow the method of maximal minors \cite{minors} and compute the 6-minors $M_a$ defined by omitting row $a$ taking a determinant and multiplying by an alternating sign (see \cite{minors} for a precise definition in terms of the $\ep$ tensor). Using Mathematica \cite{math} here an onward, $M_a$ is found to be of the form \be
	M_a=S\, K_a ~;
\ee
 the singular factor $S(x)$  is given by 
\be
	S(x) = 4\, \la_\infty\, B_3 ~;
\label{S_factor}	
\ee
 $\la_\infty,\, B_3$ were defined in (\ref{def:la_infty},\ref{def:B3}) respectively,  
  while $K_a(x)$ is given by \be
	K_a = (s_a^3,\,   s_b^1,\, s_c^2,\,  - s_a^2,\, - s_b^3,\, - s_c^1,\, 0) ~.
\label{global}
\ee

For generic values of $x \in X$ $S(x) \neq 0$ and hence $M_a(X) \neq 0$ and  the dimension of the $G$-orbit is generically 6. We confirmed this by a numerical evaluation of ${\rm rk} (Tx)$ at randomly chosen points. Since ${\rm dim}(X)=6$ we may answer the first question from the introduction section and conclude that generically in $X$ \be
 {\rm codim} (G-{\rm orbit}) = 0 ~. 
\label{G-orb}
\ee
This means that SFI is maximally effective for the triangle diagram and a discrete set of base points in $X$ space will suffice for reaching any other point through a line integral over a path which lies within a $G$-orbit.

Let us multiply the equation system (\ref{eq_set}) on the left by $K_a$. By construction $K_a$ annihilates the $Tx$ term (\ref{def:Tx}), meaning that the group action vanishes for this linear combination of generators for all $x$; hence the $K_a$ is termed the global stabilizer. Moreover $2 K_a\, c^a= \( s^2_a- s^3_a \)+ cyc. = \( x_3-x_2 \) + cyc. = 0$, namely it annihilates the $c$ term (\ref{def:c^a}) as well. This implies that $K_a J^a =0$, which we term an algebraic constraint. Indeed substituting source simplification (\ref{source_simplf}) into $J^a$ (\ref{def:J}) we have $2 K_a\, J^a = \( s^3_a\, \del_2 O_1\, I - s^2_a\, \del_3 O_1\, I \) + cyc. = \(j'_2-j'_3 \) + cyc.=0$.  

\presub {\bf Constant free invariants and the homogeneous solution}. The constant free subgroup $G_{cf}$ is defined to consist of generators such that the constant term vanishes, $c=0$. In terms of a basis of generators it consists of coefficients for linear combinations $l_a$ ($x_i$-independent) such that $l_a \, c^a =0$. In the basis (\ref{def:c^a}) this happens for $l_7=0$ and $\sum_{i=1}^6 l_a = 0$.

The $G_{cf}$ orbits are co-dimension 2, and they define 2 invariants. Denoting the differential operator appearing in (\ref{SFIeq}) by \be
D_{13} := s^1_b\, \del^3 + s^6_\infty \, \del^4 + x_5 \del^5
\label{def:D13}
\ee
we have \bea
D_{13}\, \la_\infty &=& \la_\infty \non
D_{13}\, B_3 &=& B_3~.
\label{invar}
\eea
The $D_{13}$ operator is not constant free, together with all its permutations $D_{ij}, ~ i \neq j=1,2,3$. Yet a difference of any two $D_{ij}$ operators is constant free and 
 hence $\la_\infty, B_3$ are the two $G_{cf}$ invariants. 

The homogeneous solution of the equation set (\ref{eq_set}), $I_0$, is an ingredient of the general reduction formula to a line integral. Substitution into a constant free equation shows that it is independent of directions along a $G_{cf}$-orbit and hence must depend on the $G_{cf}$ invariants, namely \be
 I_0=I_0(\la_\infty,\, B_3)~.
\ee

Substituting this into 2 (independent) non constant free equations we obtain an equation set for $I_0$ such as (\ref{SFIeq},\ref{dimeq}) \bea
	0 &=& I_0 + 2 B_3 \frac{\p I_0}{\p B_3}+2 \lambda_{\infty}\frac{\p I_0}{\p \lambda_{\infty}} \non
	0 &=& (d-6)I_0 -6 B_3 \frac{\p I_0}{\p B_3}-4 \lambda_{\infty}\frac{\p I_0}{\p \lambda_{\infty}}~.
\eea
Solving the set we find  that 
\be
	I_0 = 2 \lambda_{\infty}^{\frac{3-d}{2}} B_3^\frac{d-4}{2} \equiv \frac{2}{\sqrt{|\laminf|}}\(h^2\)^{\frac{d-4}{2}}
\label{hom}
\ee
where the multiplicative normalization was set for later convenience.

\section{Solutions}
\label{sec:solns}

\subsection{Singular locus}
\label{sec:singular}

The singular locus is defined as the locus in $X$ of non-generic $G$-orbits with sub-generic dimension \cite{locus}. On this locus there is a linear combination of SFI equations such that the differential part vanishes, namely, the equations become algebraic. If furthermore the constant term of this combination is non-zero then the diagram can be expressed as a linear combination of its descendants. Experience shows that this criterion is related to the criteria for Landau singularities \cite{Landau1959}.

Considering $S(x)$ (\ref{S_factor}), the singularity locus factor of the triangle, we see that the singularity locus consists of two components: one where $\la_\infty=0$ and one where $B_3=0$. Before proceeding to a separate discussion of each component, we describe some features which are common to both. 

At the singular locus the maximal minors are 5 dimensional and being maximal they factorize into \cite{minors} \be
 M_{ab}^i = Inv^i \, Stb_{ab}
\label{S_minor_factorization}
\ee
 $Inv^i(x)$ are the components of a 1-form in $X$ which annihilate the $G$-orbit and hence it is related to group invariants. Here it must be proportional to the gradient of the quantity which defines the locus component \be
 Inv^i = \del^i Inv
 \label{Inv_i}
 \ee 
where $Inv$ is given by either $\laminf$ or $B_3$ on the respective component. $Stb_{ab}(x)$ is a 2-form in $G$ which stabilizes (or annihilates) the point $x$. It is a 2-form since $G$ is 7d and the singular $G$-orbits are 5d. Moreover  $Stb_{ab}(x)$ defines a 2-plane in $G$ which includes in it the global stabilizer $K_a(x)$ (\ref{global}).

$Stb_{ab}$ can be computed through (\ref{S_minor_factorization},\ref{Inv_i}). Its value must be independent (mod $Inv$) of the index $i$. For the triangle we found that $Stb_{ab}$ could always be expressed as a polynomial rather than a rational function. Through degree balance we find the $x$-degree of the stabilizer to be \be
 {\rm deg}(Stb) =  {\rm deg}(M) -  {\rm deg}(Inv^i)=5-({\rm deg}(Inv)-1) = 6-{\rm deg}(Inv)
\label{stab_degree}
\ee

Next the solution at the component $Inv$ can be found by multiplying the SFI equation system indexed by $a$ (\ref{eq_set})  by $Stb_{ab}$. Now the solution must be independent of $b$ after we account for source simplification. Alternatively the solution at $Inv$ can be evaluated by Gauss elimination of derivative terms out of the equation system (for instance, implemented by Mathematica).

\presub {\bf $\laminf$ locus}. The vanishing of $\laminf$ implies that  $p_1, p_2, p_3$ are ``collinear up to a null vector'', namely that they are either collinear or that they define a degenerate plane, one where the induced metric has 0 as an eigenvalue. 

We determined the associated stabilizer $Stb_{ab}$, yet it is of degree 4 in $x$ due to (\ref{stab_degree}) and the expression was too long to be included in the paper in a useful way.

We determined the solution at the $\laminf$ locus to be \bea
 -I|_{\laminf=0} &=& \frac{1}{2\, B_3} \( \del^1 B_3\, I_1  +  \del^2 B_3\, I_2 +  \del^3 B_3\, I_3\)  = \non
			 &=& \frac{2\, x_4}{\del^1 B_3} I_1 + \frac{2\, x_5}{\del^2 B_3} I_2 + \frac{2\, x_6}{\del^3 B_3} I_3
\label{lam_soln}
\eea
where $I_i, ~i=1,2,3$ denote bubble diagrams with propagator $i$ contracted. The two lines are equal (mod $\laminf$) due to the identity (\ref{id1}), and we note that the expressions are $S_3$ symmetric, as they should be. This result has been tested successfully at the arbitrarily chosen numerical subspace $(x_4,x_5,x_6)=(7,10+2\sqrt{21},3)$. 

\presub {\bf $B_3$ locus}. The vanishing of $B_3$ implies that the tetrahedron is coplanar. 

We note that in Euclidean geometry a colinear triangle at a tetrahedron basis implies a coplanar tetrahedron. However, here $\laminf=0$ does not imply $B_3=0$. The geometrical reason is that in non-Euclidean signature $\laminf=0$ could hold when the basis triangle is contained in a null plane, and then the tetrahedron need not be degenerate. 

On this component the stabilizer is cubic in $x$, see (\ref{stab_degree}), and one form for it is \be
\( \begin{array}{c}
 x_4\, \partial^1 B_3 \\
  x_{3}\, \laminf \\
  2 \left(x_1\, x_4 \,s^4_{\infty}+x_2\, x_5\, s^5_{\infty} + x_3\, x_6\, s^6_{\infty} -  x_4\, x_5\, x_6 \right) \\
  0 \\
  x_3\, \laminf - x_5\, \partial_{1} B_3  \\
  2  x_4 \left(x_3\, x_6 - x_6\, s^6_{\infty}  - x_1\,  s^5_\infty - x_2\,  s^4_\infty \right)  - x_2\, \lambda_{\infty} \\
  - s^2_a\, \laminf \\
\end{array} \)
\ee

We find that the solution at $B_3$ is given by \bea
	I|_{B_3=0}   &=&\nonumber -\frac{2(d-3)}{(d-4)\, } \( \frac{x_4}{ \p^1 B_3} I_1 + \frac{x_5}{ \p^2 B_3} I_2 + \frac{x_6}{ \p^3 B_3} I_3 \)\\ 
 &+&\frac{2 \lambda_{\infty}}{(d-4)}\left(\frac{ x_1 T_1}{(\partial_2 B_3)(\partial_3 B_3)}+ \frac{ x_2 T_2}{(\partial_1 B_3)(\partial_3 B_3)}+\frac{ x_3 T_3}{(\partial_1 B_3)(\partial_2 B_3)}\right) ~.
\label{B3_soln}
\eea
This expression is $S_3$ symmetric, as it should be. 
The coefficient of tadpoles ($T_i$) in the above equation can be expanded in partial fractions as follows
\begin{equation}
\frac{1}{s_b^3}\left(\frac{x_5}{\partial_2 B_3}+\frac{s^4_\infty}{\partial_3 B_3}\right)=\frac{  \lambda_\infty}{(\partial_2 B_3)(\partial_3 B_3)}
\end{equation}
 and so on. The above result has been tested successfully at the arbitrarily chosen numerical values $(x_1,x_2,x_3,x_4,x_5,x_6)=(35,25,15,0,4,0)$ and $(28,60,0,44,0,-33)$.

\subsection{SFI derivation of general solution}
\label{subsec:deriv}

In this subsection we solve the SFI equation system for the triangle (\ref{eq_set}). 

According to the method of variation of the constants, once a homogenous solution, $I_0(x)$, is known we substitute into the SFI equations \be
 I(x) = I_0(x) \, \hat{I}(x)
\label{def:I_hat}
 \ee
 and we find that $\hat{I}(x)$ is given by a line integral over simpler diagrams \cite{SFI}, as will seen later for the case at hand.
 
We choose the integration curves to be the flow lines of the vector field \be
D_1 := s^1_c\, \del_2 + s^1_b\, \del^3 + x_4\,  \del^4 + x_5\,  \del^5  + x_6\,  \del^6 ~.
\label{def:D1}
\ee
This vector field is obtained by adding together the 1st and 4th rows in the SFI equation system (\ref{def:Tx}). More precisely this defines a family of curves which foliates $X$, and it can also be thought to be characteristic curves corresponding to $D_1$. 

This generator is chosen for source simplicity. Indeed  $J^1,\, J^4$, defined in (\ref{def:J}), depend only on the parameters of the $I_1\equiv O_1 I$ bubble, namely $x_2,\, x_3,\, p_1^{~2} \equiv x_4$, and the chosen linear combination will be seen to produce an especially simple source. The source is given by \be
 J_a = \( \del^2 + \del^3 \) I_1~.
\label{def:Ja}
\ee
The bubble $I_1$ is given by \be
I_1 = \frac{c_b}{\sqrt{|p_1^{~2}|}} \int_{-b_{13}}^{b_{12}} dq\, V^{\frac{d-4}{2}}
\ee
where the bubble constant is defined by  \be
c_b := i \pi^{d/2}\, \Gamma\(\frac{4-d}{2}\) ~,
\label{def:cb}
\ee
the integration limits are given by \be
 b_{13}=\frac{s^2_a}{\sqrt{|p_1^{~2}|}}  ~,	\qquad b_{12}=\frac{s^3_a}{\sqrt{|p_1^{~2}|}}
\ee
and finally the Kirchhoff-Symanzik function $V$ is given by \be
V(q) := {\vec{\hh}_1}^2 + \vec{q}^{\;2} = -\frac{\la_a}{4 p_1^{~2}} - q^2 
\ee
where the last equality holds for a space like $p_1$ and otherwise the sign of $q^2$ needs to be changed.

Putting together these ingredients we find the source to be \be
 J_a =  \frac{c_\Delta}{\sqrt{|p_1^{~2}|}} \int_{-b_{13}}^{b_{12}} dq\, V^{\frac{d-6}{2}}
\label{Ja}
\ee 
where $c_\Delta$ is the triangle constant defined in (\ref{def:cTr}), which satisfies $c_\Delta = \frac{d-4}{2}\, c_b$. We have used $0=\( \del^2 + \del^3 \) b_{13} = \( \del^2 + \del^3 \) b_{12}$ so there is no contribution from the integration limit, as well as $\( \del^2 + \del^3 \) \hh^2 =1$. All these last equalities are special for the chosen linear combination of generators and are responsible for the simple form of (\ref{Ja}) as announced above.
 
The integration curves have a geometrical interpretation: the parameter space of the diagram, $X$  (\ref{def:X}), can be identified with the parameter space of tetrahedra shown in fig. \ref{fig:tetra} hence the integration curves can be interpreted as 1-parameter families of tetrahedra. The definition of $D_1$ (\ref{def:D1}) implies \bea
 D_1 x_i &=& x_i \qquad  i=4,5,6 \non
 D_1 x_1 &=& 0 \non
 D_1 \laminf &=& 2 \laminf \non 
 D_1 B_3 &=& 2 B_3 
 \eea
where the last two equalities are implied by $D_1= D_{12}+D_{13}$ and (\ref{invar}). These relations imply that $h^2,\, \hh_2^{~2} \equiv \la_b/p_2^{~2}, \hh_3^{~2}$ are annihilated by $D_1$.

\begin{figure}[ht]
\begin{center}
\begin{tikzpicture}[scale=.5, transform shape]
   \draw[black, thick] (3,4) -- (0.8,1.9);
      \draw[black, thick] (-3,4) -- (0,4);
      \draw[black, thick,->] (3,4)--(0,4) ;
      \draw[black, thick,->] (0,-3)-- (1.5,0.5);
      \draw[black, thick] (1.5,0.5)--(3,4) ;
     \draw[black, thick,->] (-3,4)--(-1.5,0.5) ;
      \draw[black, thick] (0,-3)--(-1.5,0.5) ;
        \draw[black, thick,->] (-2.6,4)--(-1.1,0.5) ;
      \draw[black, thick] (0.235,-2.5)--(-1.1,0.5) ;
      \draw[black, thick,->] (-1.9,1.5)--(-4,0.6) ;
      \node at (-2.0,-0.5) {$p_1$};
      \node at (1.2,2.9) {$m_1$};
      \node at (0.8,1.9)[circle,fill,inner sep=2.5pt]{};
       \node at (2.0,-0.5) {$p_2$};
        \node at (0.0,4.5) {$p_3$};
\end{tikzpicture}
\caption{The flow lines of $D_1$ (\ref{def:D1}) generate families of tetrahedra where $p_1$ slides while $\hat{O}$ is fixed.}
\label{fig:tetra_fam}
  \end{center}
\end{figure}
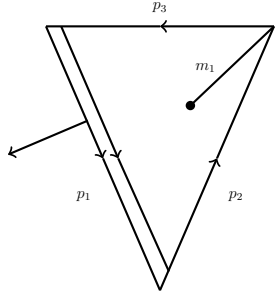

The algebraic relations above imply that the family of tetrahedra is formed by sliding $p_1$ in a direction normal to itself within the plane of $p_1, p_2, p_3$, while fixing the points $\hat{O}, O$. In this manner the $p_1, p_2, p_3$ triangle is rescaled, see fig. \ref{fig:tetra_fam}.

We need to choose a parameterization for the integration curves. Since they describe a slide of the $p_1$ edge two possibilities suggest themselves: the heights of either one of the two triangle containing $p_1$, namely either $h_1 \equiv \sqrt{|\laminf/4p_1^{~2}|}$ or $\hh_1 \equiv \sqrt{|\la_a/4p_1^{~2}|}$. In the following we shall find it convenient to use $h_1$. 

 Now we have all the necessary ingredients to integrate $I$. The $D_1$ generator defined in (\ref{def:D1}) implies the following equation for $\hat{I}$ defined in (\ref{def:I_hat})  \be
 2 I_0\, \frac{h_1}{2}  \frac{\del}{\del h_1} \hat{I}= J_a
 \ee
 where $I_0$ the homogenous solution is given in (\ref{hom}), $D_1 h_1 = h_1/2$ and the source $J_a$ is given in (\ref{def:Ja}). Performing the integration we find \be
 	\hat{I}(x) = \frac{c_\Delta}{\(h^2\)^{\frac{d-4}{2}}} \int_{\Delta_q} d^2 q\, V^{\frac{d-6}{2}} + \hat{I}(h_1=0)
 \ee
 where the integration domain $\Delta_q$ is the triangle of external momenta, see (\ref{def:Deltaq}), and $h_1=0$ was chosen as a base point (location of initial condition). This is convenient since $\hat{I}(h_1=0)=0$ as can be seen directly in the Schwinger plane, such as in subsection \ref{subsec:alpha}. Using this and restoring $I$ through (\ref{def:I_hat},\ref{hom}) we finally arrive at the same expression which appeared already in (\ref{gen_express}) by transforming the expression in Schwinger parameters into the triangle of external momenta. Thus we were able to solve the SFI  equation system for the general triangle diagram,  the expression found in this way coincides with the one obtained in subsection \ref{subsec:alpha}
 and it is the simplest expression that we have found for the diagram.
 
 \presub {\bf Comments}. We have confirmed that the general expression (\ref{gen_express}) not only satisfies the SFI equation associated with $D_1$ defined in (\ref{def:D1}) but also the one associated with $D_{13}$ defined in (\ref{def:D13}) and hence through permutations the first six equations of the SFI equation system (\ref{eq_set}). The last equation in the set is confirmed through dimensional analysis. 
 
The general expression (\ref{gen_express}) can be restricted to the singular loci and it would be interesting to compare it with the expressions (\ref{lam_soln},\ref{B3_soln}) gotten in the previous subsection.

\section{Perspective on massless triangle and magic}
\label{sec:magic}

\begin{figure}[ht]
\begin{center}
\begin{minipage}{.2\textwidth}
\begin{tikzpicture}[scale=.4, transform shape]
 \draw[black, thick] (-3.5,0) -- (3.5,0);
    \draw [thick](0,0) circle (3.5cm);
       \hspace{2.5cm} $\Longleftrightarrow$
         \node at (-8.1,3.9) {$x_1$};
         \node at (-8.1,0.4) {$x_2$};
         \node at (-8.1,-2.5) {$x_3$};
\end{tikzpicture}
\end{minipage}
\hspace{4.5cm}
\begin{minipage}{.2\textwidth}
\begin{tikzpicture}[scale=.4, transform shape]
    \draw [black,thick,domain=5.19:7.2] plot ({0}, {\x}); 
 \draw[black, dotted] (3,0) -- (0,5.19);
     \draw[black, dotted] (-3,0) -- (0,5.19);
      \draw[black, dotted] (-3,0) -- (3,0);
     \draw[black, thick] (-5,-1.3) -- (-3,0);
      \draw[black, thick] (5,-1.3) -- (3,0);
      \node at (-1,6) {$x_4$};
         \node at (-4.0,0.5) {$x_6$};
         \node at (4.0,0.5) {$x_5$};
      
\end{tikzpicture}
\end{minipage}
\caption{Magic Connection between Diameter and massless Triangle}
\label{fig:magic}
  \end{center}
\end{figure}
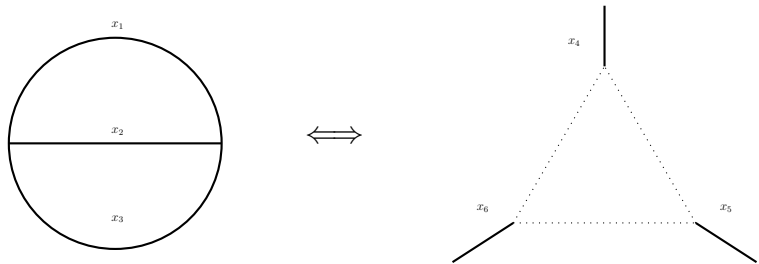

There is a connection between the diameter diagram and the massless triangle diagram upon mapping of parameters. This is known as `magic connection' in the literature \cite{magic1995} - see figure \ref{fig:magic}. We will provide a novel re-derivation of this connection by comparing the SFI equation systems for these two diagrams.

The massless triangle and the diameter are similar in that they both depend on 3 parameters and both enjoy an $S_3$ permutation symmetry. However, they differ in the diagram topology including the number of loops and the number of external legs.  Moreover, the connection includes a currently mysterious relation between the dimensions of the two diagrams.

First we tried to compare the SFI equation systems for the two diagrams, but their relation was not apparent at this level. Therefore we proceeded to compare the integral normalized by its leading singularities, defined by \be
 {\hat I} = I/I_0
\ee
where $I_0$ denotes the homogenous solution; the gradient of ${\hat I}$ is gotten by inverting the $Tx$ matrix which appears in the SFI system.

For the diameter we find \be
\p_1 \hat{I}_{D} = -\frac{d-2}{x_1 \lambda} \frac{1}{I_{0D}} \left(- x_1\, j_2 j_3 +  s^3 j_1 j_3+ s^2 j_1 j_2 \right)
\ee
 and similarly for $\p_2 \hat{I}_{D}$ and $\p_3 \hat{I}_{D}$. For the massless triangle we find \be
\p_4 \hat{I}_\Delta = \frac{2(d-3)}{x_4\, \lambda_\infty} \frac{1}{I_{0T}} \left(- x_4\, I_{1}  + s^6 I_{2} + s^5 I_{3} \right)
\ee
and similarly for $\p_5 \hat{I}_\Delta$ and $\p_6 \hat{I}_\Delta$. The respective homogenous solutions are given by \begin{eqnarray}
I_{0D}(d)&=&\lambda^{\frac{d-3}{2}}\\\nn
 I_{0 \Delta}(d)&=&\lambda_{\infty}^{\frac{3-d}{2}}(x_4 x_5 x_6)^{\frac{d-4}{2}}
\end{eqnarray}
while the tadpole and bubble sources are given by 
\begin{eqnarray}
j_\mu(\mu ; d)&=&i \pi^\frac{d}{2} \Gamma(2-\frac{d}{2})\mu^{(\frac{d}{2}-1)} \label{magic_sources} \\
I_{B i}(\mu_i; d)&=&\frac{i^{1-d}\pi^{\frac{d}{2}}\Gamma (2-\frac{d}{2})\Gamma^{2}(\frac{d}{2}-1)}{\Gamma(d-2)}\mu_i^{(\frac{d}{2}-2)} \nonumber
\end{eqnarray}

After substituting for these we get 
\bea
\p_1 \hat{I}_{D} &=& -\frac{(d-2)\, c_T^2}{x_1\, \lambda} \frac{1}{\lambda^{\frac{d-3}{2}} } \left(- x_1\, (x_2\, x_3)^{\frac{d-2}{2}}  + s^2\, (x_1\, x_2)^{\frac{d-2}{2}} + s^3\,  (x_1\, x_3)^{\frac{d-2}{2}}  \right) \non
\p_4 \hat{I}_\Delta &=& \frac{2\, (d-3) c_B\,}{x_4\, \lambda_\infty}\,  \lambda^{\frac{d-3}{2}}  \left(- x_4\, (x_5\, x_6)^{-\frac{d-4}{2}}  + s^5 (x_4\, x_6)^{-\frac{d-4}{2}}+ s^6 (x_4\, x_6)^{-\frac{d-4}{2}}  \right)
\label{I_hat_DT}
\eea
where $c_T, c_B$ are  the tadpole and bubble constants, respectively, which depend only on $d$ and can be read off (\ref{magic_sources}).

Now it is observed that the two equations are similar under the straightforward mapping \be
 x_i \leftrightarrow x_{i+3}\equiv p_i^2,  ~~~ i=1,2,3~.
 \ee
However, this is not enough. In order to match the powers of $\lambda$ we must have \be
 d_D + d_\Delta=6 ~,
 \label{d_magic}
\ee
namely the respective dimensions $d_D$ and $d_\Delta$ must change. (\ref{d_magic}) implies $d_D-2 \leftrightarrow 4-d_\Delta$ and hence the expressions within parenthesis in (\ref{I_hat_DT}) match as well. 

Finally after multiplication by an $x$-independent factor we obtain the magic connection \begin{equation}
I_D(x_1,\, x_2,\, x_3 ;\, d)=i^{1-d} \pi^{\frac{3d}{2}-3}\frac{\Gamma(3-d)}{\Gamma(\frac{d}{2})}(x_1 x_2 x_3)^{\frac{d}{2}-1}I_\Delta (\{p_i^2=x_i\}_{i=1,2,3} ;\, 6-d) ~.
\end{equation}
This result matches exactly with the relation discovered in \cite{magic1995}. The original derivation used the Mellin-Barnes representation while we provide a novel re-derivation through SFI. Unfortunately the current derivation does not motivate the dimension relation (\ref{d_magic}) but at least it makes clear how the correspondence works given this relation.

\section{Summary and discussion}
\label{sec:summ}

In this paper we have analyzed the triangle Feynman integral through the Symmetries of Feynman Integrals (SFI) method. The SFI analysis stresses the relation of any diagram with simpler diagrams obtained through edge contraction, diagrams which can be termed descendants. For the triangle the descendant diagrams are three different bubble diagrams, see fig. \ref{fig:sources}.

We proceed to summarize the paper's results. The SFI equation system was determined and presented in a simple basis in (\ref{eq_set}).  We studied the geometry of parameter space and found that the SFI method is maximally effective here as the co-dimension of the G-orbit is 0 (\ref{G-orb}). The singular locus was found to consist of two components where either the  Heron / K\"all\'en invariant $\laminf$ or the Tartaglia / Baikov polynomial $B_3$ vanish (\ref{S_factor}). At these components the triangle was evaluated as a linear combination of descendant bubble diagrams (\ref{lam_soln},\ref{B3_soln}). 

The general solution was derived in subsection \ref{subsec:deriv}, arriving at an expression (\ref{gen_express}) in terms of an integral over the triangle of external momenta. This expression was already essentially known since \cite{DavydychevDelbourgo1997} and it can be derived directly by transforming the alpha (Schwinger) parameter representation as described in subsection \ref{subsec:alpha}.  It is the simplest expression that we know for the general triangle, it can be decomposed (split) into a sum of 6 terms (\ref{IsumF}) and it must be equal to the known expression in terms of Appell functions, as discussed in the third paragraph below (\ref{dilog}). 

This list of results answers the first three questions posed in the introduction; now we address the fourth. Sum decomposition was known to originate from a split of the integration domain and SFI does not add to this perspective. The magic connection was discussed through the SFI perspective in section \ref{sec:magic} but the transformation of dimensions remains mysterious.  

\presub {\bf Discussion}. Following \cite{DavydychevDelbourgo1997} we stressed the underlying tetrahedron geometry. The analytic expressions contain numerous appearances of the quantities $(\laminf, B_3)$ defined in (\ref{def:la_infty},\ref{def:B3}) which are instances of Cayley-Menger (CM) determinants that express the volume of a simplex in terms of the length squares of its edges. Instances of such appearances include the tetrahedron height (\ref{def:V0_and_h}) and the singular locus (\ref{S_factor}). Much of the geometry of a simplex such as the tetrahedron can be expressed in terms of CM determinants, and this perspective forms the basis of a mathematical field known as \emph{Distance Geometry} which has applications to GPS navigation and MRI tomography, see e.g. \cite{Liberti_Lavor2015}. Hence we realize that Distance Geometry plays a role in the evaluation of the triangle diagram and likely also in more general diagrams. 

\subsection*{Acknowledgments}
 
We would like to thank Ruth Shir and Amit Schiller for many useful discussions and the workshop ``The Mathematics of Linear Relations between Feynman Integrals'' MITP, Mainz for hospitality while this work was in progress. For further hospitality B.K. would like to thank Harald Ita (Freiburg university), while S. M. would like to thank TIFR, Mumbai and International Solvay Institutes, Brussels. We would like to thank Phillip Burda for participation in early stages of this work.
 
This research was supported by the ``Quantum Universe'' I-CORE program of the Israeli Planning and Budgeting Committee.

B.K. dedicates this paper to Neta.

\appendix

\section{The $n$-simplex}
\label{sec:n_simplex}

This appendix contains some higher dimensional generalization of the discussion of the tetrahedron geometry in subsection \ref{sec:def}. This should be useful for more involved diagrams including 1-loop diagrams with more legs.

\presub {\bf The $n$-dimensional simplex and its Cayley-Menger determinant}. The $n$-dimensional simplex $\Delta_n$ is the polytope defined by $n+1$ points (or vertices) $u_0, u_1,\dots, u_n$ \cite{simplex,wiki_simplex}. The standard simplex is defined by the $n+1$ standard basis vectors in $\IR^{n+1}$. The low-dimensional simplices are the point, interval, triangle and tetrahedron for $n=0,1,2$ and $3$, respectively.

Given an $n$-dimensional simplex, the Cayley-Menger (CM) determinant \cite{wiki_Cayley-Menger} 
 is defined by the $(n+2)*(n+2)$ determinant \be
C_n = \det \left[ \begin{array}{cccccc}
	0 		& d_{01}^2 &  d_{02}^2 & \dots &  d_{0n}^2 & 1 \\
	d_{10}^2  & 0		 &  d_{12}^2 & \dots &  d_{1n}^2 & 1 \\
	\vdots  	& \vdots  	& \vdots 	    & 	\ddots	& \vdots 	& \vdots \\
	d_{n0}^2  & d_{n1}^2 &  d_{n2}^2   & \dots 	   &  0 & 1 \\
	1	& 1	& 1	&	\dots & 1	& 0 \\
\end{array} \right]
\label{def:Cayley}
\ee
where \be
 d_{ij}^2 :=\( \vec{u}_i -\vec{u}_j \)^2
 \label{def:dij}
 \ee
and in the last definition the vector notation $\vec{u}_i$ stresses the vector nature of $u_i$. Equivalently \be
 C_n = (-)^{n+1}\, 2^n \det \{ s_{ij} \}_{i,j=1,\dots,n}
\label{def:Cayley2}
\ee
where $s_{ij} = \(d_{0i}^2+d_{0j}^2-d_{ij}^2 \) /2$. This definition is in terms of a smaller, $n *n$ determinant, but it hides the $S_n$ permutation symmetry. 

The Cayley-Menger determinant is related to the squared volume of $\Delta_n$ through the normalization \be
  {\rm Vol}^2(\Delta_n) = \frac{(-)^{n+1}}{n!^2\, 2^n} \, C_n ~.
 \label{Vol_norm}
  \ee
It is manifestly symmetric under the $S_{n+1}$ permutations of the vertices.

The distances $d_{ij}^2$ are known to fix the embedding of the system of points into Euclidean space as long as some positivity conditions on CM determinants of sub-simplices hold (the conditions include $d_{ij}^2 \ge 0$ for each edge and the triangle inequalities for each triangle). Moreover, we believe that any set of squared distances fixes an embedding into a pseudo-Euclidean space of free signature, and it is not clear to us whether this generalization appears already in the literature.  
    
In low dimensions we have \bea
 C_0  &=& -1 \non
 C_1 &=& 2\, d_{01}^2 \non
 C_2 &=& \la\(d_{01}^2,d_{02}^2,d_{12}^2\) \non
 C_3 &=& -2 B_3
\eea
and \bea
  {\rm Vol}^2(\Delta_0) &=& 1  \non
  {\rm Vol}^2(\Delta_1) &=& d_{01}^2  \non
  {\rm Vol}^2(\Delta_2) &=& -\la/16   \non
 {\rm Vol}^2(\Delta_3) &=& -B_3/144   ~.
\eea

\presub {\bf Identity for derivatives of a CM determinant}. We have found useful identities for the derivative of $C_n$ which generalize (\ref{id1}-\ref{id3}) to arbitrary dimensions.\footnote{
We thank Nadav Drukker for a useful and enjoyable discussion of this subject.}
 From the definition (\ref{def:Cayley}) and the determinant derivative formula (Jacobi's formula \cite{wiki_Jacobis_formula})
 we have \be
\frac{\del}{\del d_{01}^2} C = - 2\, C_{12}
\ee
where $C_{ij}$ is the minor obtained from $C$ by deleting row $i$ and column $j$ and taking the determinant, and symmetry implies $C_{12}=C_{21}$.  On the other hand the minors of any matrix $M$ (not necessarily symmetric) satisfy \be
 M_{11}\, M_{22} - M_{12}\, M_{21} = M\, M_{1212}
 \ee
 where $M_{1212}$ is the minor obtained by deleting rows $1,2$ and columns $1,2$. This can be proven by expanding the determinant with respect to rows $1,2$ (row expansion is known as the Laplace expansion  \cite{Laplace_expansion,wiki_Laplace_expansion}). In addition, we verified this formula for $2*2$ and $3*3$ matrices.  
Combining these two equations we obtain \be
  C_{11}\, C_{22} - \frac{1}{4} \( \frac{\del}{\del d_{01}^2} C \)^2 = C\, C_{1212}~.
\label{id1n}
 \ee 
Noting that $C_{11}, C_{22}, C_{1212}$ are all Cayley-Menger determinants of sub-simplices, this identity expresses the derivative of $C$ in terms of CM determinants. Permutation symmetry immediately generalizes the discussion to a derivative of $C$ with respect to any $d_{ij}^2$.

For $n=2$ dimensions (\ref{id1n}) becomes \be
 4\, x_1\, x_2\, - 4\, (s^3)^2 = -\la
 \ee 
 where $\la=\la(x_1,x_2,x_3)$ and $s^3=-\del \la/(4\del x_3)$ is an $s$ variable (\ref{def:s}); for $n=3$ it reproduces (\ref{id1}).

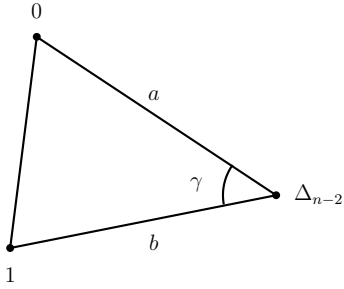
\begin{figure}[ht]
\begin{center}
\begin{tikzpicture}[scale=.7, transform shape]

 \draw[black, thick] (3,0) -- (-1.5,3);
    
    \draw[black, thick] (3,0) -- (-2,-1);
    \draw[black, thick] (-2,-1) -- (-1.5,3);
    \node at (3.8,0.0) {$\Delta_{n-2}$};
    \node at (1.5,0.2) {$\gamma$};
    \node at (0.7,1.9) {$a$};
     \node at (0.7,-0.9) {$b$};
     \node at (-2,-1.5) {$1$};
     \node at (-1.5,3.5) {$0$};
     \node at (-2,-1)[circle,fill,inner sep=1.5pt]{};
     \node at (3,0)[circle,fill,inner sep=1.5pt]{};
     \node at (-1.5,3)[circle,fill,inner sep=1.5pt]{};
      \draw [black,thick ,domain=190:145] plot ({3+1*cos(\x)}, {1*sin(\x)});
\end{tikzpicture}
\caption{A projection of $\Delta_n$ over $\Delta_{n-2}$, a sub-simplex which does not include the $0,1$ vertices.  $a,b$ are the lengths of the shown edges and $\gamma$ is the angle between them. This figure illustrates the geometrical interpretation of derivatives of CM determinants as discussed in the text.}
\label{fig:n-2_proj}
  \end{center}
\end{figure}

\presub {\bf Geometric interpretation}. We proceed to offer a geometric interpretation for the identity (\ref{id1n}). For this purpose we shall use fig. \ref{fig:n-2_proj} which shows the simplex $\Delta_n$ and the $01$ edge. $\Delta_n$ is projected over the hyperplane containing the simplex $\Delta_{n-2}$ that consists of all points other than $0,1$, and hence $\Delta_{n-2}$ collapses into a point and the projection is planar.

Denoting by $\Delta^1_{n-1}$ the $(n-1)$-simplex obtained by deleting vertex $1$ from $\Delta_n$ we have \be
	{\rm Vol}\, \Delta^1_{n-1} = \frac{1}{n-1}\, a\, {\rm Vol}\, \Delta_{n-2}
  \label{Vn-1}
\ee
 and similarly by exchanging $1, a$ into $0, b$ respectively. In addition we have \be
  {\rm Vol}\, \Delta_n = \frac{2}{n(n-1)}\, {\rm Vol}\, \Delta_2\, {\rm Vol}\, \Delta_{n-2}
 \label{Vn-2}
  \ee
 where $\Delta_2$ is the triangle shown in the figure. Henceforth we shall assume that the plane of the triangle $\Delta_2$ is Euclidean and hence its area  is of course \be
  {\rm Vol}\, \Delta_2 = \frac{1}{2}\, a\, b\, \sin \gamma 
\label{V2}
 \ee
 (for a non-Euclidean triangle the $\sin$ function should be appropriately generalized into a hyperbolic function). 
 Putting together these ingredients we have \bea
  {\rm Vol}^2 \, \Delta_n\, {\rm Vol}^2 \, \Delta_{n-2} &=& \frac{4}{n^2(n-1)^2}\, {\rm Vol}^2 \, \Delta_2\, {\rm Vol}^4 \Delta_{n-2} =\non
  	&=& \frac{a^2\, b^2 \sin^2 \gamma}{n^2(n-1)^2} \, {\rm Vol}^4 \, \Delta_{n-2} = \non
	&=&   \frac{(n-1)^2}{n^2}\, {\rm Vol}^2 \, \Delta^0_{n-1}\, {\rm Vol}^2 \, \Delta^1_{n-1}\,  \sin^2 \gamma ~.
  \eea
The first equality uses (\ref{Vn-2}), the second (\ref{V2}) and the third substituted $a^2, b^2$ from  (\ref{Vn-1}).

Changing normalizations into CM determinants through (\ref{Vol_norm}) we obtain \be
 C_n\, C_{n-2} = C_{n-1}^0\, C_{n-1}^1 \sin^2 \gamma ~.
 \ee
Now (\ref{id1n}) implies \be
 \frac{1}{4} \( \frac{\del}{\del d_{01}^2} C \)^2 = \sqrt{C_{n-1}^0\, C_{n-1}^1} \cos^2 \gamma
 \label{id3n}
\ee
and the identity (\ref{id1n}) reduces to $\sin^2 \gamma + \cos^2 \gamma =1$. Equation (\ref{id3n}) expresses a CM derivative in terms of the volumes of the two relevant $(n-1)$-simplices and the angle between them (in fact, once the areas are viewed as tensors, this is their inner product) and it generalizes (\ref{id3}) to arbitrary dimensions.

Another identity can be obtained by noticing that upon choosing an $(n-1)$ simplex, say the one associated with vertices $1,2,\dots,n$ its volume is equal to the sum of the projections of all other $(n-1)$-simplices onto it. This implies \be
 \( \frac{\del}{\del d_{01}^2} + \frac{\del}{\del d_{02}^2} + \dots + \frac{\del}{\del d_{0n}^2} \) C = -2\, C^0_{n-1}
 \ee
 where we have used (\ref{id3n}). This identity generalizes (\ref{id2}) from $n=3$ to all dimensions and provides it with a geometrical interpretation. In particular, for $n=2$ it reads \be
 \( \del_1 + \del_2 \) \la = -4 x_3
 \ee
 which holds since $\del_i \la = -4\, s^i$ for $i=1,2,3$.

\bibliographystyle{JHEP}
\bibliography{triangle_bib}

\end{document}